\documentclass[%
 aip,
 amsmath,amssymb,
 reprint,%
]{revtex4-1}

\usepackage{graphicx}
\usepackage{dcolumn}
\usepackage{bm}

\usepackage[utf8]{inputenc}
\usepackage[T1]{fontenc}
\usepackage{mathptmx}
\usepackage{etoolbox}
\usepackage{wrapfig}

\makeatletter
\def\@email#1#2{%
 \endgroup
 \patchcmd{\titleblock@produce}
  {\frontmatter@RRAPformat}
  {\frontmatter@RRAPformat{\produce@RRAP{*#1\href{mailto:#2}{#2}}}\frontmatter@RRAPformat}
  {}{}
}%
\makeatother
\begin{document}

\preprint{AIP/123-QED}

\title{Resonance Cascades and Number Theory}
\author{Oleksandr V. Marchukov}
 \affiliation{Technische Universit\"{a}t Darmstadt, Institut f\"{u}r Angewandte Physik, Hochschulstra{\ss}e 4a, 64289 Darmstadt, Germany.}
 \affiliation{Institute of Photonics, Leibniz University Hannover, Nienburger Straße 17, D-30167 Hannover, Germany}
\author{Maxim Olshanii}%
 \email{maxim.olchanyi@umb.edu.}
\affiliation{ 
Department of Physics, University of Massachusetts Boston, Boston Massachusetts 02125, USA
}%

\date{\today}

\begin{abstract}
In this article, we are interested in situations where the existence of a
contiguous cascade of quantum resonant transitions is predicated on the validity
of a particular statement in number theory. The setting is a tailored one-atom one-dimensional potential with a prescribed 
spectrum, under a weak periodic perturbation. The former is, by now, an experimental reality [D. Cassettari, G. Mussardo and A. Trombettoni, PNAS Nexus {\bf 2}, pgac279 (2022)]. 
As a case study, we look at the
following trivial statement: ``Any power of $3$ is an integer.'' Consequently, we
``test'' this statement in a numerical experiment where we demonstrate an unimpeded
upward mobility along an equidistant, $\ln(3)$-spaced subsequence of the energy
levels of a potential with a log-natural spectrum, under a frequency $\ln(3)$
time-periodic perturbation. We further show that when we ``remove'' $9$ from the set of 
integers---by excluding the corresponding energy level from the spectrum---the cascade halts 
abruptly.    
\end{abstract}

\maketitle

\section{Introduction}
The links between number theory and quantum physics can not yet be labeled as ``numerous.'' The majority of the text-book quantum 
problems constitute the solvable ones. The latter possesses a regular spectrum that appears to be disparate from the quasi-random nature of the number-theoretical objects. The notable exception is the quantum factorization algorithm \cite{shor1994_124}; it does not relate on properties of the prime numbers, and, hence, it does not contain any quasi-random elements.  Quantum-chaotic systems are, by their nature, closer to the number theory entities. The most prominent example of such a connection is the suspected interpretation of the Riemann zeros as eigenenergies of an $xp$ Hamiltonian with yet unknown boundary conditions \cite{berry1999_236}.       

The situation changed when it became apparent that some artificial quantum objects can be, instead, experimentally tailored to acquire number-theoretical properties   \cite{cassettari2023_279}. (See also the proposal \cite{latorre2013_13026245}.)
In this article, we show how to create a quantum system, similar to the one in \cite{cassettari2023_279}, where the existence of a
contiguous cascade of quantum resonant transitions, in a one-body one-dimensional system is predicated on the validity
of a particular statement in number theory. The empirical relevance of our exploits is backed by the recent experiments  \cite{cassettari2023_279} on creating quantum atomic potentials with a tailored spectrum. Our work derives its inspiration from a host of previous proposals for analogue quantum number-theoretical machines; there, the focus is mainly on the problem of prime factorization  (see \cite{gleisberg2018_035009} and references therein and \cite{mussardo2023_10757}). 

Our article is a part of a broader cycle of works set to explore the number-theory-inspired effects in quantum systems and how best to make them prominent. In particular it can be regarded as an improvement over as similar scheme suggested  
\cite{marchukov2024_241013988}; the latter suffered from the localization of the atomic population, requiring some ad hoc improvements. In the current article, this problem is resolved by means of choosing a more optimal spatial profile of the perturbation, \eqref{V_cumulative}. 
Also, the same article studies fidelity of the individual two-state transitions under a periodic perturbation, in a class of number-theory potentials.  An experimental proof of the Goldbach conjecture \cite{dudley_Number_Theory1969} (p. 147) belongs to a group of more distant goals of this program (see, also, \cite{prudencio2014_13036476}). A scheme enhanced by a quantum speed-up was proposed in \cite{marchukov2024_240400517}. An alternative scheme involving two atoms can be found in \cite{marchukov2024_241013988}. The article \cite{carmona2024_240719898}
establishes a relationship between Mersenne numbers \cite{dudley_Number_Theory1969} (p. 60) and ground states in certain spin lattices.  Finally \cite{ramesh2024_241102436} utilizes a Brahmagupta identity \cite{stillwell} and an identity due to H\'{e}l\`{e}ne Perrin   \cite{jackson2024_062} to explain systematic degeneracies in a particular rectangular quantum billiard.  

%
\section{The model and the strategy}
We will focus on a potential whose spectrum is logarithms of natural numbers. Powers of a given integer for an equidistant ladder, with a possibility of igniting a resonance
cascade of transitions under an external periodic drive. Existence of a resonance cascade will serve as a quantum-analogue ``proof'' that positive integer powers of an integer belong to a set of natural  numbers. We will also attempt to simulate a universe where one of the powers is excluded from the natural set: in that case, our analogue machine will indeed halt.

%
\subsection{The Hamiltonian}
Consider a one-body potential $U^{\text{L}}(x)$ whose lowest $N$ bound state energies are logarithms of natural numbers:
\begin{align}
\begin{split}
&
E_{n} = U_{0} \ln n
\\
&
n=1,\, 2,\, 3,\,\ldots,\, N_{\text{b}}
\end{split}
\label{log_n}
\,\,.
\end{align}
Methods of designing tailored potentials are described in \cite{cassettari2023_279}. 
The approximate  classical counterpart of the $\ln(n)$-potential is a logarithmic potential: 
\begin{align}
U_{\text{cl.}}(x) = U_{0} \ln(\sqrt{\frac{2}{\pi}} \frac{x}{a})
\,\,;
\label{log_n_potential_classical}
\end{align}
here \[a\, \equiv \hbar/\sqrt{mU_{0}}\,\,.\] Indeed, let us treat the neighboring energy level differences as classical frequencies, 
$E_{n+1}-E_{n} \approx \hbar\omega(E \approx E_{n} \approx E_{n+1})$, arriving at 
\[
\omega(E) = \frac{U_{0}}{\hbar} e^{-\frac{E}{U_{0}}}
\,\,.
\] 
Assume that the classical potential we are looking for, $U_{\text{cl.}}(x)$ is even, $U_{\text{cl.}}(-x)=U_{\text{cl.}}(+x)$. Also, assume that the potential is infinitely deep, so that $U_{\text{cl.}}(x) \stackrel{x\to 0}{\longrightarrow} -\infty$. Using a known result
\cite{book_landau_mechanics} (\S 12) that  
allows one to relate the dependence of the classical frequency on energy, $\omega(E)$ to the potential $U_{\text{cl.}}(x)$ that governs this motion, one gets the following expression for the inverse of $U_{\text{cl.}}(x)$ for $x>0$:
\[
x(U_{\text{cl.}}) = \int_{-\infty}^{U_{\text{cl.}}} \frac{dE'}{\omega(E')\sqrt{2m(U_{\text{cl.}}-E')}} 
= \sqrt{\frac{\pi}{2}} \frac{\hbar}{\sqrt{m U_{0}}} e^{\frac{U_{\text{cl.}}}{U_{0}}}
\,\,.
\]
Inverting this expression with respect to $U_{\text{cl.}}$ leads us to \eqref{log_n_potential_classical}.

\subsection{The tight-binding model}
Consider a monochromatic time-dependent 
perturbation 
\begin{align}
V(x) \cos(\Omega t)
\,\,,
\label{log_n_perturbation}
\end{align}
with
\begin{align}
&
\Omega = \frac{1}{\hbar}U_{0} \ln(\tilde{n})
\,\,,
\label{log_n_perturbation_frequency}
\end{align}
for some $\tilde{n}$ that optimizes the speed of a particular time-dependent process, for some profile $V(x)$. Namely, we want to ensure that the level population can propagate freely through any subset of the unperturbed energy levels \eqref{log_n} that is simply connected via
resonant (with respect to the perturbation \eqref{log_n_perturbation_frequency}) transitions. Naively, one may choose a parametric
excitation (where the coupling constant $U_{0}$ acquires a small time-dependent correction) as the most experimentally relevant. 
However an investigation in \cite{marchukov2024_241013988} showed that the rapid decay of the matrix elements of the perturbation as a function of energy results in a localization. Below, we undergo a more systematic search for the optimal spatial profile $V(x)$.  

Similarly to \cite{marchukov2024_241013988}, we will use a resonant approximation to predict and control the population dynamics.   
Our Schr\"{o}dinger equation reads:
\begin{align}
\begin{split}
&
i\hbar \frac{\partial}{\partial t}\Psi = \hat{H}_{0}\Psi + V(x) \cos(\Omega t) \Psi
\\
&
\Psi = \Psi(x,\,t)
\\
&
\Psi(x,\,0)=\Psi_{0}(x)
\\
&
 \hat{H}_{0} = -\frac{\hbar^2}{2m} \frac{\partial^2}{\partial x^2} + U^{\text{L}}(x)
 \end{split}
\,\,.
\label{schrodinger_time-dependent}
\end{align}
Let us formally decompose $\Psi(x,\,t)$ onto a Floquet series with the nature of the coefficients $\chi_{l}(x,\,t)$, to 
be discussed later in the text:
\[
\Psi(x,\,t) = \sum_{l=-\infty}^{+\infty} \chi_{l}(x,\,t)\exp[+i l\Omega t]
\,\, .
\] Let us now substitute the above series to the Schr\"{o}dinger equation \eqref{schrodinger_time-dependent}. We get:
\begin{align*}
&
\sum_{l=-\infty}^{+\infty} 
i\hbar \frac{\partial}{\partial t}\chi_{l}(x,\,t) \exp[+i l\Omega t]
\\
\quad
&
= 
\sum_{l=-\infty}^{+\infty} 
 (\hat{H}_{0}+\hbar l \Omega )\chi_{l}(x,\,t) \exp[+i l\Omega t]
 +  
 \\
&
\quad\quad \sum_{l=-\infty}^{+\infty} 
 \frac{1}{2} V(x) \chi_{l}(x,\,t) \exp[+i (l+1)\Omega t]
 +
 \\
&
\quad\quad\quad
 \sum_{l=-\infty}^{+\infty} 
 \frac{1}{2} V(x) \chi_{l}(x,\,t) \exp[+i (l-1)\Omega t]
\\
&
\sum_{l=-\infty}^{+\infty} \chi_{l}(x,\,0)=\Psi_{0}(x)
\,\,.
\end{align*}
This system of equations for $\chi_{l}(x,\,t)$ is greatly overdetermined. However, if one (a)  
requires that in the resulting Schr\"{o}dinger equation prefactors in front of each 
oscillating exponent  $\exp[+i l'\Omega t]$ be balanced individually
and (b) reserves the $l=0$ harmonic to encode the initial state, the resulting system is well determined: 
\begin{align*}
&
i\hbar \frac{\partial}{\partial t}\chi_{l}(x,\,t)  = (\hat{H}_{0}+\hbar l \Omega )\chi_{l}(x,\,t)  
+  
\\
&
\quad \frac{1}{2}V(x) \left( \chi_{l-1}(x,\,t) +\chi_{l+1}(x,\,t) \right)
\\
&
\chi_{l}(x,\,0)=\delta_{l,0}\Psi_{0}(x) 
\,\,.
\end{align*}

As the next step, we decompose the above equation onto a basis of the eigenstates of $\hat{H}_{0}$:
\begin{align*}
&
\chi_{l}(x,\,t) = \sum_{n=1}^{\infty} \chi_{l,n}(t) \psi_{n}(x)
\\
&
\hat{H}_{0} \psi_{n} = E_{n}  \psi_{n}
\,\,.
\end{align*}
For the coefficients $\chi_{l,n}(t)$, 
we get 
\begin{align*}
&
i\hbar \frac{\partial}{\partial t}\chi_{l,n}(t) =  (E_{n}+\hbar l \Omega )\chi_{l,n}(t) +
\\
&
\quad  
\frac{1}{2}\sum_{n'=1}^{\infty}\langle n | V | n' \rangle (\chi_{l-1,n'}(t)+\chi_{l+1,n'})(t)
\\
&
\chi_{l,n}(0)=\delta_{l,0} \langle n | \Psi_{0} \rangle 
\,\,.
\end{align*}
Notice that the unperturbed manifold that contains the initial state is massively degenerate. Indeed, since $\hbar \Omega = U_{0} \ln(\tilde{n})$ and $E_{n} = U_{0} \ln(n)$, all the 
Floquet quasi-energies with indices $n=\tilde{n}^m,\,l=-m$ have zero Floquet  energy 
$\mathcal{E}_{l,n}\equiv E_{n} + \hbar l \Omega$:
\[
\mathcal{E}_{-m,\,\tilde{n}^m} = E_{\tilde{n}^m} + \hbar (-m) \Omega = 0
\,\,.
\] 

If the perturbation is sufficiently weak, one can restrict oneself to the zero order of the degenerate perturbation theory where the matrix elements $\langle n | V | n' \rangle$ of the perturbation leading to the states outside of the $\mathcal{E} =0$ degenerate manifold 
(where $n=\tilde{n}^{-l}$)
are neglected. There one arrives at a  tight-binding model:
\begin{align}
i\hbar\frac{\partial}{\partial t} \phi_{m} = -J_{m+1,m} = \phi_{m+1} -J_{m,\,m-1} \phi_{m-1} 
\,\,,
\label{tight_binding}
\end{align}
where
\begin{align}
&
\phi_{m}(t) \equiv  \chi_{\tilde{n}^m,-m}
\nonumber
\\
&
J_{m,\,m-1}= J_{m-1,\,m} = \langle \tilde{n}^{m} | V | \tilde{n}^{m-1}   \rangle
\label{hopping}
\,\,.
\end{align}

In a \emph{constant-hopping} tight-binding model ($J_{m}=J=\text{const}$), population density propagates ballistically. For an initial state $\phi_{m}(0) = \delta_{m,0}$, the r.m.s. and absolute values of the position $m$ will grow as 
\begin{align*}
&
\sqrt{\langle m^2 \rangle} = \sqrt{2} J t
\\
&
\langle |m| \rangle \approx 1.27 J t
\,\,.
\end{align*}
Recall also that the maximal speed of sound in this model is $c=2J$. 

A semi-infinite tight-binding model, 
\begin{align*}
&
m \ge 0
\\
&
\phi_{-1} = 0
\,\,,
\end{align*}
is closer to the problem at hand. There one finds
\begin{align}
&
\sqrt{\langle m^2 \rangle} \stackrel{t\gg\hbar/J}{\approx} -0.67+1.71 J t
\label{wall_rms}
\\
&
\langle m \rangle \stackrel{t\gg\hbar/J}{\approx}  -0.76+1.68 J t
\label{wall_abs}
\,\,.
\end{align}
Hence, if one finds a way to ensure a relative uniformity of the hopping coefficients \eqref{hopping}, one may expect to find a linear in time growth of energy. Recall that for the cascade states, their energies are \[E_{n=\tilde{n}^{m}} = U_{0} \ln(\tilde{n}) m\,\,. \] 

\emph{Our goal} is to find a perturbation $V(x)$ that produces a set of relatively homogeneous set of the hopping constants $J_{m}$ \eqref{hopping}.

\section{Finding the optimal perturbation profile, $V(x)$}
In this Section, our goal is to design a perturbation that generates a tight-binding model with relatively homogeneous hopping coefficients \eqref{hopping}. The technical difficulty is that while the semiclassical expressions for the eigenstate wavefunctions are valid in the area of parameters we are interested in, the relevant changes in quantum numbers are too great to support the standard estimate for the transition matrix elements as classical Fourier transforms
\cite{landau_quantum___matrix_elements_as_fourier}. This case was worked out in \cite{landau_quantum___matrix_elements_difficult}. 

For a single $\delta$-peak, 
\begin{align}
&
V^{(\delta)}(x|\,g,\,x_{0}) = g \delta(x-x_{0})
\label{V_delta_single}
\,\,,
\end{align}
the WKB prediction \cite{landau_quantum___matrix_elements_difficult} for the matrix elements leads to the following estimate
\begin{align}
\begin{split}
&
V^{(\delta)}_{n_{1},n_{2}} \sim \frac{g}{a} \, \frac{1}{\sqrt{n_{1}n_{2}}} \times
\\
&
\quad
\left\{
\begin{array}{lll}
1 & \text{for} & n_{1} > n_{\rm min}(x_{0}) \text{ and }  n_{2} > n_{\rm min}(x_{0})
\\
0 & \text{otherwise} &
\end{array}
\right\}
\end{split}
\label{WKB_delta_single}
\,\,,
\end{align}
where
\begin{align*}
&
n_{\rm min}(x_{0}) = \sqrt{\frac{2}{\pi}} \,\frac{x_{0}}{a}
\end{align*}
is the lowest quantum number such that the corresponding eigenstate has $x_{0}$ inside the classically allowed region.  In \eqref{WKB_delta_single}, 
slow, logarithmic with $n$ contributions are neglected. 

The result \eqref{WKB_delta_single} can be understood as follows. For $n_{1} \sim n_{2} \sim n$, the matrix element of the 
$\delta$-perturbation \eqref{V_delta_single} can be estimated as $g$ times the classical probability density (see e.g.\ 
\cite{olshanii_book_back_of_envelope_II}, Eq.\ 5.32) 
\[
\rho_{\text{class.}}(x) = \frac{\omega(E)}{\pi v(x\,|\,\,E)}
\] 
at $x_{0}$, where $E\sim E_{n_{1}} \sim E_{n_{2}} \sim U_{0} \ln(n)$, $\omega(E)$ is the classical frequency, and 
$v(x\,|\,\,E) = \sqrt{2(E-V(x))/m}$ is the magnitude of the classical velocity. Notice that beyond the classical aphelion,  where
$V(x) = E$ or $x= \sqrt{\pi/2} \, a \exp(E/U_{0})$, the density vanishes. Furthermore, since the spatial variation of the density is 
logarithmically slow, in an estimate, it can be replaced by a constant. All in all, we get 
\begin{align*}
\begin{split}
&
V^{(\delta)}_{n_{1},n_{2}} \sim \frac{g}{a} \, \frac{1}{n} \times
\left\{
\begin{array}{lll}
1 & \text{for} & n > n_{\rm min}(x_{0}) 
\\
0 & \text{otherwise} &
\end{array}
\right\}
\end{split}
\\
&
\text{with}
\\
&
n \sim n_{1} \sim n_{2}
\,\,;
\end{align*}
this estimate is consistent with a more accurate expression \eqref{WKB_delta_single}.
 
The \eqref{WKB_delta_single} scaling inspires the following form for the perturbation:
\begin{align}
&
V^{(\delta, \, \text{array})}(x) = \sum_{i=1}^{\infty} V^{(\delta)}(x|\,g,\,(x_{0})_{i})
\label{V_delta_cumulative}
\,\,,
\end{align}
with 
\[
(x_{0})_{i} = \Delta x \times i
\,\,.
\]
For the potential \eqref{V_delta_cumulative}, we get, for the hopping constants \eqref{hopping}
\[
J_{m} \sim \frac{g}{\sqrt{\tilde{n}}\Delta x} = \text{const}(m)
\,\,,
\]
the result we were hoping for.

In our numerical calculations, we replace the $\delta$-function by a finite width Gaussian:
\begin{align}
&
V_{\sigma}(x|\,g,\,x_{0}) = g D_{\sigma}(x-x_{0})
\label{V_single}
\,\,,
\end{align}
with 
\begin{align}
&
D_{\sigma}(x) = \frac{1}{\sqrt{2\pi}\sigma} e^{-\frac{x^2}{2\sigma^2}}
\label{D_single}
\,\,.
\end{align}
Accordingly, the infinite array of $\delta$-potentials is replaced by a finite array of Gaussian perturbation:
\begin{align}
&
V^{\text{(array})}(x) = \sum_{i=1}^{i_{\rm max}} V_{\sigma}(x|\,g,\,(x_{0})_{i})
\label{V_cumulative}
\,\,,
\end{align}
with 
\[
(x_{0})_{i} = \Delta x \times i
\,\,,
\]
and 
with 
\[
\sigma_{i} = \Delta \sigma \times i
\,\,.
\]

\section{Numerical Results}

Our numerical experiment consists of two parts. First, in order to construct the logarithmic potential we use the QM-SUSY approach described in \cite{cassettari2023_279}. We solve the associated Riccatti equations numerically, using standard SciPy library methods \cite{virtanen2020_261}. Second, we solve the one-dimensional Schr\"{o}dinger equation with the obtained potential (and perturbation given in the text) using fast Fourier transform-based spectral method for spatial discretization and a fourth-order symplectic integrator 
scheme for the time evolution. We use 16384 points for both spatial and temporal grids, to ensure that the relative error between the lowest numerical eigenvalues of the potential and the correct eigenvalues is not larger that $10^{-6}$.

We used the perturbation \eqref{V_cumulative} with $g=0.01$, $\Delta x = 2.82$, $\Delta \sigma = 0.25$, and $i_{\rm max} = 10$. The resulting numerical resonance cascade hopping constants read:
\begin{align*}
\begin{split}
&
\{
J_{1,0},\,J_{2,1},\,J_{3,2},\,J_{4,3}
\}
=
\\
&
\quad
\{
0.00027,
0.00049, 
0.00024,  
0.00011
\} \times U_{0}
\end{split}
\,\,.
\end{align*}

In the first experiment, all natural numbers $1 \le n \le 120$ were present 
in the $\ln(n)$ spectrum. The atomic population was able to  travel freely among all five cascade levels, all the way to the highest available 

\begin{figure}[!h]
\begin{center}
\includegraphics[width=.9\textwidth]{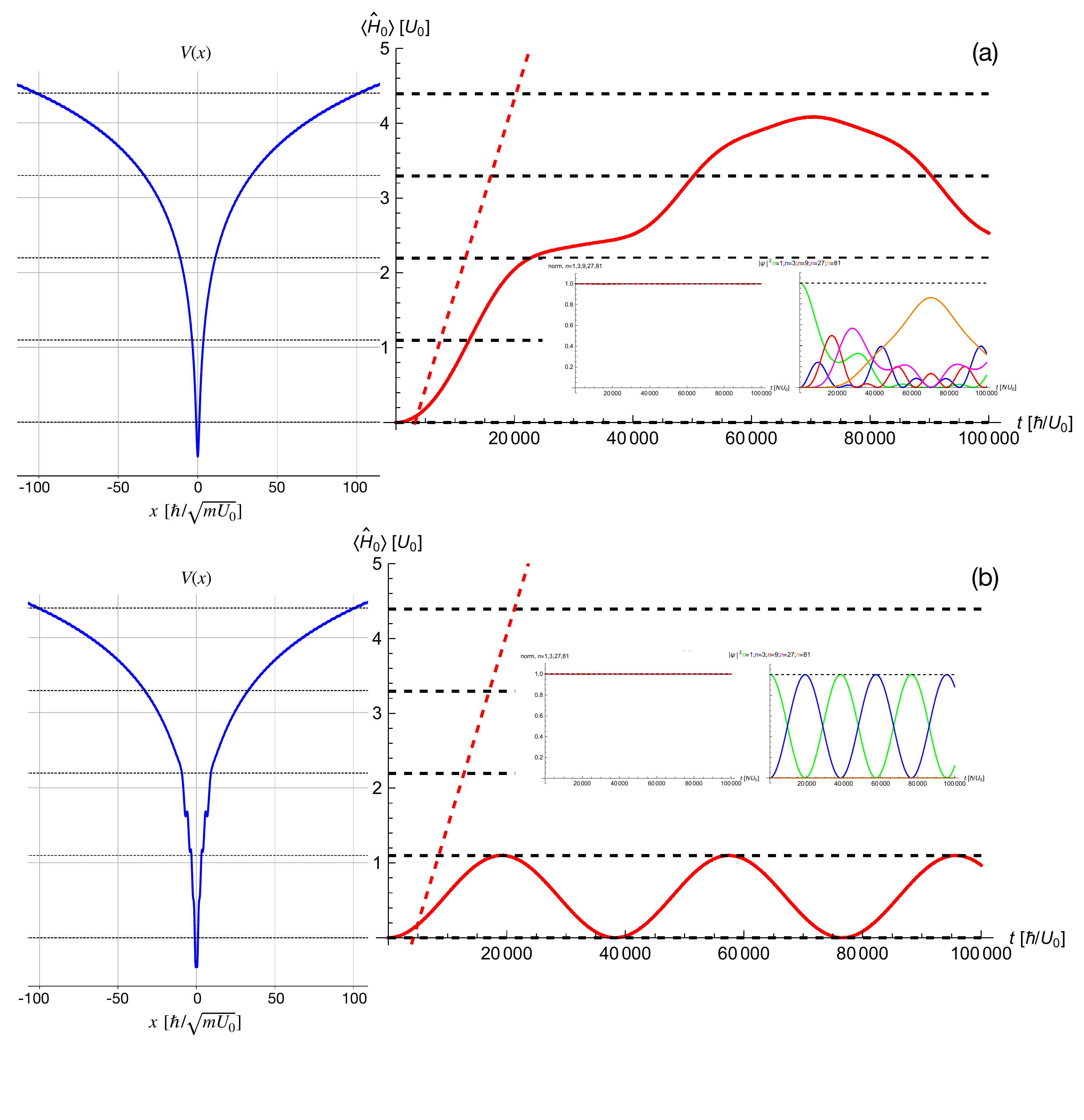}
\end{center} 
\caption{
\textbf{A ``test'' for closeness of a subset of natural numbers under multiplication. }
In this numerical experiment, we want to see if an atom can travel from the ground state ($n=1$) to every single available power of $3$ state ($n=3^{m}$), 
under  an action of a weak monochromatic perturbation of a frequency $\Omega = U_{0} \ln(3)$, 
in an unperturbed potential with spectrum $E_{n} = U_{0} \ln(n)$ where the indices 
$1 \le n \le 120$ form a  subset of the set of natural numbers. The spatial dependence of the perturbation potential was designed specifically to ensure an approximate uniformity of the transition matrix elements across the spectrum (see text). 
The main plot shows energy as a function of time (solid red). 
The straight line (dashed red) is a linear fit \eqref{wall_abs} to the analogous time dependence for a semi-infinite constant-hopping lattice of the resonant states, $n=1,\,3,,\,9,\,\ldots,\,+\infty$ with the hopping constant 
being equal to the average of the corresponding hopping coefficients of the full model. The insert to the left shows the unperturbed  potential. The lower left insert reflects the the total 
population of the resonant cascade states $n=1,\,3,,\,9,\,27,\,81$ (red), as compared to unity (dashed black). 
The lower right insert shows the populations of the individual cascade states, $n=1$ (green), $n=3$ (green), $n=9$ (red), $n=27$ (magenta), $n=1$ (orange). (a) All natural numbers $1 \le n \le 120$ are present 
in the $\ln(n)$ spectrum.  (b) We design a new potential where the energy level $n=9$ is excluded.
}
\label{f:together}
\end{figure}

We also designed a new potential where the energy level $n=9$ was excluded. This created bottleneck in the middle of the 
resonance cascade. In this case, the atomic population does not propagate further than $n=3$, undergoing Rabi oscillations between $n=1$ and $n=3$.
Our numerical results are presented in Fig.~\ref{f:together}.

\section{Conclusion, Context, and Outlook}

In this article, we attempted to create an atomic system where 
\begin{itemize}
\item[(a)] the movement of the population between the energy levels under a periodic perturbation
occurs exclusively via exact resonances while at the same time the atomic population does explore \emph{all} phase space reachable via the resonant transitions.
\item[(b)] a possibility for an unlimited growth of energy  is predicated by the truthfulness of a number-theoretical statement;   such growth can be disrupted by a single violation of the statement. 
\end{itemize}
In the example we looked at, the unlimited growth of energy required the closeness of the set of natural numbers (\emph{including} zero) under multiplication. This was achieved using an atom in a logarithmic potential perturbed by a comb of narrow Gaussian peaks (in space) with periodically oscillating magnitudes (in time).

In the future, we would like to apply the scheme of the current article to a set of sums of two squares of natural numbers (including zero). This set is also known to be closed under multiplication, thanks to 
the  Diophantus-Brahmagupta-Fibonacci identity
\cite{dudley_Number_Theory1969} (Lemma 1, p.\ 142). 

%

%

\begin{acknowledgments}
We are grateful to Donatella Cassettari, Andrea Trombettoni, Giuseppe Mussardo, and Joanna Ruhl for their valuable comments. A significant portion of this work was produced during the thematic trimester on ``Quantum Many-Body Systems Out-of-Equilibrium'', at the  Institut Henri Poincaré (Paris): MO is immeasurably grateful to the organizers of the trimester, Rosario Fazio, Thierry Giamarchi, Anna Minguzzi, and Patrizia Vignolo,  for an opportunity to be a part of it.  

OVM acknowledges the support by the DLR German Aerospace Center with funds provided by the Federal Ministry for Economic Affairs and Energy (BMWi) under Grant No. 50WM1957 and No. 50WM2250E. MO was supported by the NSF Grant No.~PHY-1912542. The authors would like to thank the Institut Henri Poincar\'{e} (UAR 839 CNRS-Sorbonne Université) and the LabEx CARMIN (ANR-10-LABX-59-01) for their support.
- and the Institut Henri Poincar\'{e} (UAR 839 CNRS-Sorbonne Université) and LabEx CARMIN (ANR-10-LABX-59-01).
\end{acknowledgments}

\section*{Author Declarations}
\subsection*{Conflict of interest}
The authors have no conflicts to disclose.

\section*{Data Availability Statement}
The data that support the findings of this study are available from the corresponding author upon reasonable request.

\section*{Bibliography}

\bibliography{Bethe_ansatz_v059,Nonlinear_PDEs_and_SUSY_v053}

\end{document}